\documentstyle[prd,epsf,aps]{revtex}
\input{psfig}
\newcommand{\pom}{I\!\! P}

\begin{document}
\draft
\title{Diffractive dijet production at $\sqrt s=630$ and $1800$ GeV at the 
Fermilab Tevatron}
\maketitle
\font\eightit=cmti8
\def\r#1{\ignorespaces $^{#1}$}
\hfilneg
\begin{center}

\vglue 1em
\centerline{\large The CDF Collaboration}

\vglue 1em
\centerline{\large (Submitted to Physical Review Letters)}
\vglue 1em

\begin{sloppypar}

\noindent
T.~Affolder,\r {23} H.~Akimoto,\r {45}
A.~Akopian,\r {37} M.~G.~Albrow,\r {11} P.~Amaral,\r 8  
D.~Amidei,\r {25} K.~Anikeev,\r {24} J.~Antos,\r 1 
G.~Apollinari,\r {11} T.~Arisawa,\r {45}
% ...
\end{sloppypar}
\end{center}
\vskip .026in
\begin{center}
\r 1  {\eightit Institute of Physics, Academia Sinica, Taipei, Taiwan 11529, 
Republic of China} \\
\r 2  {\eightit Argonne National Laboratory, Argonne, Illinois 60439} \\
\r 3  {\eightit Istituto Nazionale di Fisica Nucleare, University of Bologna,
I-40127 Bologna, Italy} \\
\r 4  {\eightit Brandeis University, Waltham, Massachusetts 02254} \\
\r 5  {\eightit University of California at Davis, Davis, California  95616} \\
\r 6  {\eightit University of California at Los Angeles, Los 
Angeles, California  90024} \\  
\r 7  {\eightit Instituto de Fisica de Cantabria, CSIC-University of Cantabria, 
39005 Santander, Spain} \\
\r 8  {\eightit Enrico Fermi Institute, University of Chicago, Chicago, 
Illinois 60637} \\
\r 9  {\eightit Joint Institute for Nuclear Research, RU-141980 Dubna, Russia}
\\
\r {10} {\eightit Duke University, Durham, North Carolina  27708} \\
\r {11} {\eightit Fermi National Accelerator Laboratory, Batavia, Illinois 
60510} \\
\r {12} {\eightit University of Florida, Gainesville, Florida  32611} \\
\r {13} {\eightit Laboratori Nazionali di Frascati, Istituto Nazionale di Fisica
               Nucleare, I-00044 Frascati, Italy} \\
\r {14} {\eightit University of Geneva, CH-1211 Geneva 4, Switzerland} \\
\r {15} {\eightit Glasgow University, Glasgow G12 8QQ, United Kingdom}\\
\r {16} {\eightit Harvard University, Cambridge, Massachusetts 02138} \\
\r {17} {\eightit Hiroshima University, Higashi-Hiroshima 724, Japan} \\
\r {18} {\eightit University of Illinois, Urbana, Illinois 61801} \\
\r {19} {\eightit The Johns Hopkins University, Baltimore, Maryland 21218} \\
\r {20} {\eightit Institut f\"{u}r Experimentelle Kernphysik, 
Universit\"{a}t Karlsruhe, 76128 Karlsruhe, Germany} \\
\r {21} {\eightit Center for High Energy Physics: Kyungpook National
University, Taegu 702-701; Seoul National University, Seoul 151-742; and
SungKyunKwan University, Suwon 440-746; Korea} \\
\r {22} {\eightit High Energy Accelerator Research Organization (KEK), Tsukuba, 
Ibaraki 305, Japan} \\
\r {23} {\eightit Ernest Orlando Lawrence Berkeley National Laboratory, 
Berkeley, California 94720} \\
\r {24} {\eightit Massachusetts Institute of Technology, Cambridge,
Massachusetts  02139} \\   
\r {25} {\eightit University of Michigan, Ann Arbor, Michigan 48109} \\
\r {26} {\eightit Michigan State University, East Lansing, Michigan  48824} \\
\r {27} {\eightit University of New Mexico, Albuquerque, New Mexico 87131} \\
\r {28} {\eightit The Ohio State University, Columbus, Ohio  43210} \\
\r {29} {\eightit Osaka City University, Osaka 588, Japan} \\
\r {30} {\eightit University of Oxford, Oxford OX1 3RH, United Kingdom} \\
\r {31} {\eightit Universita di Padova, Istituto Nazionale di Fisica 
          Nucleare, Sezione di Padova, I-35131 Padova, Italy} \\
\r {32} {\eightit University of Pennsylvania, Philadelphia, 
        Pennsylvania 19104} \\   
\r {33} {\eightit Istituto Nazionale di Fisica Nucleare, University and Scuola
               Normale Superiore of Pisa, I-56100 Pisa, Italy} \\
\r {34} {\eightit University of Pittsburgh, Pittsburgh, Pennsylvania 15260} \\
\r {35} {\eightit Purdue University, West Lafayette, Indiana 47907} \\
\r {36} {\eightit University of Rochester, Rochester, New York 14627} \\
\r {37} {\eightit Rockefeller University, New York, New York 10021} \\
\r {38} {\eightit Rutgers University, Piscataway, New Jersey 08855} \\
\r {39} {\eightit Texas A\&M University, College Station, Texas 77843} \\
\r {40} {\eightit Texas Tech University, Lubbock, Texas 79409} \\
\r {41} {\eightit Institute of Particle Physics, University of Toronto, Toronto
M5S 1A7, Canada} \\
\r {42} {\eightit Istituto Nazionale di Fisica Nucleare, University of Trieste/
Udine, Italy} \\
\r {43} {\eightit University of Tsukuba, Tsukuba, Ibaraki 305, Japan} \\
\r {44} {\eightit Tufts University, Medford, Massachusetts 02155} \\
\r {45} {\eightit Waseda University, Tokyo 169, Japan} \\
\r {46} {\eightit University of Wisconsin, Madison, Wisconsin 53706} \\
\r {47} {\eightit Yale University, New Haven, Connecticut 06520} \\
\r {(\ast)} {\eightit Now at Carnegie Mellon University, Pittsburgh,
Pennsylvania  15213} \\
\r {(\ast\ast)} {\eightit Now at Northwestern University, Evanston, Illinois 
60208} \\
\r {(\ast\ast\ast)} {\eightit Now at University of California, Santa Barbara, CA
93106}
\end{center}
\vglue 0.5in
\centerline{\large Abstract}

\begin{abstract}
We report a measurement of the diffractive structure function $F_{jj}^D$ 
of the antiproton obtained from a study of dijet events produced in 
association with a leading antiproton in $\bar pp$ collisions at 
$\sqrt s=630$ GeV at the Fermilab Tevatron. The ratio of $F_{jj}^D$ 
at $\sqrt s=630$ GeV to $F_{jj}^D$ obtained from a similar measurement at 
$\sqrt s=1800$ GeV is compared with expectations from QCD factorization and 
with theoretical predictions. We also report a measurement of the $\xi$ 
($x$-Pomeron) and $\beta$ ($x$ of parton in Pomeron) dependence of $F_{jj}^D$
at $\sqrt s=1800$ GeV. In the region $0.035<\xi<0.095$, $|t|<1$ GeV$^2$ 
and $\beta<0.5$,  $F_{jj}^D(\beta,\xi)$ is found to be of the form 
$\beta^{-1.0\pm 0.1} \xi^{-0.9\pm 0.1}$, which obeys $\beta$-$\xi$ 
factorization. 
%The measured $\xi^{-0.9\pm0.1}$ dependence 
%indicates that dijet production within our kinematical range 
%is dominated by Pomeron exchange. 
\end{abstract}


In a previous Letter~\cite{CDF_JJ_RP_1800}, we reported a 
measurement of the diffractive structure function of the antiproton 
extracted from events with two jets produced in association with a 
leading (high momentum) antiproton in $\bar pp$ collisions 
at $\sqrt s=1800$ GeV at the Fermilab Tevatron.
Conceptually, diffractive jet production may be thought of as 
a two-step process, $\bar p+p\rightarrow [\bar p'+\pom]+p\rightarrow 
\bar p'+Jet_1+Jet_2+X$, where a Pomeron~\cite{Regge}, $\pom$, 
emitted by the $\bar p$ interacts with the proton to produce the jets.
In this picture, the structure function of the Pomeron in terms of $\beta$ 
(momentum fraction of $\pom$ carried by its struck parton) at a
given value of $\xi$ (momentum fraction of $\bar p$ carried by $\pom$) 
is directly related to the ``diffractive structure function" of the 
antiproton in terms of the familiar Bjorken variable $x$ through the relation  
$x=\beta\xi$. 
A question of interest is whether the Pomeron,
which in QCD is a color-singlet construct of (anti)quarks and gluons 
carrying the quantum numbers of the vacuum, 
has a unique partonic structure. 
This question was addressed in our previous Letter~\cite{CDF_JJ_RP_1800}
by comparing our measured Pomeron structure
with a prediction based on diffractive parton 
densities extracted by the H1 Collaboration from 
a QCD analysis of deep inelastic scattering (DIS) data 
obtained at the DESY $ep$ collider HERA.
A disagreement was found,
expressed mainly as a suppression of ${\cal{O}}(10)$ of the overall 
normalization of our data relative to the prediction, 
indicating a severe breakdown of QCD factorization in diffractive processes. 

The suppression of the $\bar p$ diffractive structure function at the Tevatron 
relative to that at HERA 
is generally attributed to low-$x$ partons in the proton interacting
with the final state leading $\bar p$ and thus spoiling the diffractive 
signature of the event~\cite{R,GLM,softcolor,KKMR}. Consequently, the 
diffractive structure function is expected to increase as the $\bar pp$ 
c.m.s. energy, $\sqrt{s}$, decreases.   
An indirect indication of such an effect may 
have been seen in our measurement of 
the diffractive structure function of the proton in events with 
a leading antiproton, whose presence in the event 
restricts the maximum energy available 
in the diffractive subsystem~\cite{CDF_DPE}.
In this Letter, we report a measurement of 
the diffractive structure function of the antiproton at $\sqrt s=630$ GeV
and test QCD factorization by comparing it with our measurement 
at $\sqrt s=1800$ GeV. In addition, we examine the question of 
$\beta$-$\xi$ factorizarion within the differential form of the
diffractive structure function at $\sqrt s=$1800 GeV.
Diffractive dijet production in $\bar pp$ collisions at 
$\sqrt s=$ 630 GeV has been studied by the UA8 Collaboration at the 
CERN $Sp\bar pS$ collider~\cite{UA8}, but the results reported were not 
presented in terms of a normalized Pomeron structure function which 
could be directly compared with our 1800 GeV measurement.

The present study is identical to our previous 
diffractive dijet study in the experimental setup used for 
data collection and in methodology~\cite{CDF_JJ_RP_1800}. 
Briefly, a Roman Pot Spectrometer (RPS) was employed 
to trigger the CDF detector on leading antiprotons from 
single diffractive (SD) events, $\bar pp\rightarrow \bar p' X$. 
%with no other trigger requirement imposed. 
In the off-line analysis, the fractional momentum loss $\xi$  of the $\bar p$
and the 4-momentum transfer squared $t$ were determined 
with resolutions $\delta\xi=\pm 1.5\times 10^{-3}$ 
and $\delta t=\pm 0.02$ GeV$^2$
using RPS information and the event vertex. 
The RPS acceptance at $\sqrt s=$ 630 GeV is very similar to that at 1800 GeV
at the same $\xi$ and for $t$ scaled down by a factor of $(1800/630)^2=8.2$. 
The data were collected in 1995-96 (Run 1C)
with the Tevatron running at $\sqrt s=630$ GeV at an average instantaneous  
luminosity of $\sim 1.3\times 10^{30}$ cm$^{-2}$ sec$^{-1}$.
After applying off-line cuts 
requiring a reconstructed track in the RPS, 
a single reconstructed vertex in the CDF detector  
within $|z_{vtx}|<60$~cm, and a  
multiplicity of less than 5 in a forward beam-beam counter (BBC) 
array on the downstream side 
of the $\bar p$ beam, BBC$_{\bar p}$,  we obtained 
184327 SD events in the region $0.035<\xi<0.095$ and $|t|<0.2$ GeV$^2$. 
BBC$_{\bar p}$ is one of two 16 channel scintillation counter arrays
which covers the region $-5.9<\eta<-3.2$~\cite{CDFDET}, where $\eta$ is 
the pseudorapidity of a particle defined in terms of the polar angle $\theta$ 
as $\eta=-\ln \tan\frac{\theta}{2}$ (the other BBC array, BBC$_p$, covers the 
region $3.2<\eta<5.9$). 
The BBC$_{\bar p}$ multiplicity cut 
is applied to further reject overlap events 
that pass the single vertex requirement. The overlap events, 
consisting of a non-diffractive (ND) event superimposed on a SD, are
due to multiple interactions occurring in the same beam-beam bunch crossing.
The fraction of SD events 
rejected by this cut is $\approx 2.1\%$, and the ND background 
in the remaining SD sample is $\approx 2.9\%$.

Using the above inclusive SD data set, 
we selected a SD dijet sample containing 1186 SD 
events with at least two jets of corrected transverse energy 
$E_T^{jet}>7$ GeV. Similarly, a ND dijet sample of 104793 events 
was selected from a data set of 2.5 million events collected with a 
trigger requiring a BBC$_p$-BBC$_{\bar p}$ coincidence.
The $E_T^{jet}$ was defined as the sum of the
calorimeter $E_T\equiv E\sin\theta$ 
within a cone of radius 0.7 in $\eta$-$\phi$ 
space~\cite{clustering}, where $\phi$ is the azimuthal angle.
The jet energy correction included 
subtraction of an average underlying event $E_T$ of 0.5 (0.9) GeV 
for SD (ND) events. These values were determined 
experimentally, separately for SD and ND events, 
from the $\sum E_T$ of calorimeter tower energy measured 
within a randomly chosen $\eta$-$\phi$ cone of radius 0.7 in events of the 
inclusive SD and ND data samples.

The diffractive dijet sample contains a residual 
$(6.4\pm 2.2)\%$ overlap events, 
as determined from an analysis of the BBC 
multiplicity distributions. Each diffractive data distribution presented 
below is corrected for the overlap background by subtracting the corresponding  
ND distribution normalized to the overlap fraction.
Another correction is 
due to the single vertex selection requirement.
In addition to rejecting events from multiple interactions, this 
requirement also rejects single interaction events 
with multiple vertices caused by reconstruction ambiguities 
in high multiplicity events. 
From an analysis of the 
BBC and forward calorimeter tower multiplicities, the single vertex cut 
efficiency (fraction of single interaction events retained 
by the single vertex cut) was 
determined to be $(88.0\pm 1.2)\%$. 

Figure~1 presents the dijet mean $E_T$ and mean $\eta$ distributions,
$E_T^*=(E_T^{jet1}+E_T^{jet2})/2$ and 
$\eta^*=(\eta^{jet1}+\eta^{jet2})/2$,
for the SD (points) and ND (histograms) event samples.
As in the 1800 GeV case, the SD  $E_T^*$ distribution is somewhat 
steeper than the ND, and the SD
$\eta^*$ is boosted towards the proton direction (positive $\eta^*$). 
These features indicate that the $x$ dependence of the diffractive structure 
function of the antiproton is steeper than that of the ND, as discussed 
further below.
 
The $\bar p$ diffractive structure function is evaluated following 
the procedure described in our previous Letter~\cite{CDF_JJ_RP_1800}.
The fraction $x$ of the momentum of the $\bar p$ carried by the struck parton
is determined from the $E_T$ and $\eta$ of the jets using the equation
$x=\frac{1}{\sqrt{s}}\sum_{i=1}^nE_T^ie^{-\eta^i}$.
The sum is carried out over the two leading jets plus the next highest 
$E_T$ jet, if there is one with $E_T>5$ GeV.
In leading order QCD, the ratio $R(x)$ of the SD to ND rates 
is equal to the ratio of the  SD to ND structure functions of the $\bar p$.
The diffractive structure function may therefore be obtained by multiplying 
$R(x)$ by the known ND structure function.
The absolute normalization of the SD dijet sample is obtained 
by scaling the dijet event rate to that of the inclusive diffractive sample 
and using for the latter the previously measured inclusive cross 
section~\cite{CDFD}.
The normalization of the ND dijet sample is determined from our 
previously measured 
$39.9\pm 1.2$ mb cross section of the BBC trigger.

Figure~2 shows the ratio $\tilde{R}(x)$ of the number of 
SD dijet events, corrected for RPS acceptance, to the number of 
ND dijets, after normalizing both SD and ND samples 
to correspond to the same luminosity (black points). 
For comparison, $\tilde{R}(x)$ is also shown 
for our 1800 GeV data (open circles) within the same kinematic region.
The tilde over $R$ indicates integration over 
$(t,\xi,E_T^{jet})$ for SD and $E_T^{jet}$ for ND events.
The integration is carried out over the regions of $|t|<0.2$ GeV$^2$, 
$0.035<\xi<0.095$ and $E_T^{jet1,2}>7$ GeV. 
To minimize possible normalization shifts between the two data 
sets resulting from the different underlying event levels at the two 
energies, or from the influence of the third jet on the $E_T$ values of the 
leading jets, a cut was imposed on the average dijet transverse energy 
requiring $E^*_T>10$ GeV.
The ratios $\tilde{R}(x)$ exhibit similar $x$ dependence at the two 
energies, but the 630 GeV points lie systematically above the 1800 GeV ones.
A discrepancy between the two ratios would be evidence for a 
breakdown of factorization. The observed effect is 
quantified below after discussion of the 
relevant systematic errors.

As mentioned above, $R(x)$ represents the ratio of the 
diffractive to ND parton densities of the antiproton,
as viewe by dijet production. The associated 
structure functions can be written as $F_{jj}(x)=x[g(x)+\frac{4}{9}q(x)]$,
where $g(x)$ is the gluon and $q(x)$ the quark density, which is 
multiplied by $\frac{4}{9}$ to account for color factors.
The diffractive structure function $\tilde{F}^D_{jj}(\beta)$ 
is obtained by multiplying $\tilde {R}(x)$ by the ND 
structure function $F^{ND}_{jj}(x)$ and changing variables from 
$x$ to $\beta$ using the relation $x=\beta \xi$. 
The ND structure function was evaluated using GRV98LO parton 
densities~\cite{pdf}.

Figure~3 shows $\tilde{F}^D_{jj}(\beta)$, expressed per unit $\xi$, 
for the 630 GeV (black points) and 1800 GeV (open circles) data.
The curves are fits of the form $\tilde{F}^D_{jj}(\beta)=B(\beta/0.3)^{-n}$ 
in the range $0.1<\beta<0.5$. The value $\beta=0.1$ corresponds to 
the limit $x_{min}=4\times 10^{-3}$ imposed on the 
630 GeV data to guarantee full detector acceptance for the
dijet system from diffractive events associated with the  
lowest $\xi$ value of 0.035; the upper limit of 
$\beta=0.5$ is the value below which the measured $\tilde{F}^D_{jj}(\beta)$ 
at 1800 GeV was found to have a power law behaviour~\cite{CDF_JJ_RP_1800}. 
The fits yield $B=0.262\pm 0.030$ ($0.193\pm 0.005$) and $n=1.4\pm 0.2$ 
($1.23\pm 0.04$) at $\sqrt s=$ 630 (1800) GeV, 
where the quoted uncertainties are 
statistical. Within these uncertainties, the $n$ parameters are 
consistent with being equal at 
the two energies. Fitting the 630 GeV data using the parameter 
$n$ measured at 1800 GeV yields $B_{630}=0.255\pm 0.029$.

The ratio of the 630 to 1800 GeV $B$ parameters is 
$R_B=1.3\pm 0.2\,{\rm (stat)}^{+0.4}_{-0.3}\,{\rm (syst)}$.
The systematic error is due to two sources. 
The first source is 
the uncertainty in the relative normalization between the two energies. 
This is taken to be the sum 
in quadrature of a $\pm 4.5\%$ uncertainty in the ratio  of the BBC trigger 
cross sections at the two energies, and a +0.4 signed uncertainty 
%representing the difference in $R_B$ 
resulting from the difference between the experimentally measured inclusive SD 
cross section at $\sqrt s=1800$ GeV within our ($\xi,t$) region, 
$\sigma^{exp}=0.57\pm 0.03\,{\rm (stat)}$ mb (obtained from Eqs.~(3) and (4) 
in~\cite{CDFD}), and the cross section 
derived from a global fit to SD cross sections, 
$\sigma^{fit}=0.40\pm 0.04\,{\rm (syst)}$~\cite{GM}.
The second source of systematic uncertainty is a signed uncertainty of $-0.3$,
representing the difference in $R_B$ resulting 
from using only two or up to four instead of three jets in 
an event in determining the values of $x$-Bjorken.
Other possible systematic uncertainties, for example those associated with 
jet energy scale, are less important, as they tend to cancel out in the 
measurement of SD to ND ratios, and to an even higher degree in the 
measurement of the ratio of ratios.  

A deviation of $R_B$ from unity quantifies the  breakdown of
factorization. The measured value of $R_B$ is consistent with 
factorization, but is also consistent with the prediction
$R_B^{ren}=(1800^2/630^2)^{2[\alpha(0)-1]}=1.55$ 
of the renormalized Pomeron flux model~\cite{R}, evaluated using  
$\alpha(0)=1.104$~\cite{GM} for the Pomeron intercept, and with the value  
of 1.8 expected in the rapidity gap survival model of~\cite{KKMR}.
 
To further characterize the diffractive structure function, we have measured
its dependence on $\beta$ and $\xi$ (Fig.~4) 
using the higher statistics 1800 GeV data sample
of events with $E_T^{jet1,2}>7$ GeV. 
In the region $\beta<0.5$ and $0.035<\xi<0.095$, the data are well represented 
by the factorizable form
\begin{equation}
F_{jj}^D(\beta,\xi)=C\cdot \beta^{-n}\cdot \xi^{-m}
\end{equation}
The circle-points in Fig.~4a [Fig.~4b] are the values $n$ 
[$F_{jj}^D(\beta,\xi)|_{\beta=0.1}$] of a fit of Eq.~(1) 
to the data with $\beta<0.5$ 
within the indicated $\xi$-bin.  
A straight line one parameter fit to the points in Fig.~4a and a fit of the 
form $\xi^{-m}$ to those in Fig.~4b
yield $n=1.0\pm 0.1$ and $m=0.9\pm 0.1$, respectively, where the errors 
are mainly due to the systematic uncertainty associated with the 
measurement of the $\beta$ of the struck parton of the antiproton.
The observed $\xi$ dependence is steeper than that 
of the inclusive SD data sample, which is also shown in Fig.~4b (triangles).
In Regge theory, the rather flat shape of the inclusive $dN/d\xi$ 
distribution results from the superposition of a Pomeron exchange contribution,
which has a $\xi^{-\alpha(0)}\approx \xi^{-1.1}$ dependence, and a 
Reggeon exchange contribution, which enters with an effective pion 
trajectory~\cite{GM} and is $\sim \xi$. The 
measured $\xi^{-0.9\pm 0.1}$ dependence indicates that dijet production is 
dominated by Pomeron exchange.

A similarly steep $\xi$ dependence is exhibited by the 
$F_2^{D(3)}(\beta,\xi,Q^2)$ structure function extracted from 
diffractive DIS at HERA in the region $\xi<0.04$~\cite{H1,ZEUS}.
Our result of $m\approx 1$ shows that a predominantly Pomeron-like 
behaviour, which is generally expected in the small $\xi$ region explored 
by HERA, is also realized at moderately large $\xi$ values in diffractive 
dijet production at the Tevatron.
Such behaviour is predicted by models in which the structure of the generic 
Pomeron is effectively built from the non-diffractive parton densities by 
two exchanges, one at the high $Q^2$ scale 
of the hard scattering and the other at the hadron mass scale of 
${\cal{O}}(1\;{\rm GeV}^2)$~\cite{softcolor,KKMR,KG_JPG}. 

In summary, we have measured the diffractive structure function 
of the antiproton from dijet production 
in $\bar pp$ collisions at $\sqrt s=$ 630 GeV and 
compare it with that measured previously at $\sqrt s=1800$ GeV 
to test factorization. We find shape agreement between the two structure 
functions and a normalization ratio of 
$1.3\pm 0.2\,{\rm {\rm (stat)}}^{+0.4}_{-0.3}\,{\rm (syst)}$.
Within the quoted uncertainties, this ratio is compatible with the 
factorization expectation of unity, but is also compatible  
with the phenomenological predictions of 1.55 and 1.8 of 
the Pomeron flux renormalization~\cite{R} 
and gap survival probability models~\cite{KKMR}, respectively.
We have also measured the $\beta$ and $\xi$ dependence of the 
diffractive structure function at $\sqrt s=1800$ GeV 
and find that it obeys $\beta$-$\xi$ factorization for $\beta<0.5$.
The observed $\xi^{-0.9\pm 0.1}$ dependence shows 
that Pomeron-like behaviour extends to 
moderately high $\xi$ values in diffractive dijet production,
which is mainly sensitive to the gluon 
content of the diffractive structure function.
Such behaviour is expected in models in which 
the Pomeron emerges from the quark-gluon sea as a combination of two 
partonic exchanges, one on a hard scale that produces the dijet system 
and the other on a soft scale that neutralizes the color flow and forms 
the rapidity gap~\cite{softcolor,KKMR,KG_JPG}. 

     We thank the Fermilab staff and the technical staffs of the
participating institutions for their vital contributions.  This work was
supported by the U.S. Department of Energy and National Science Foundation;
the Italian Istituto Nazionale di Fisica Nucleare; the Ministry of Education,
Science, Sports and Culture of Japan; the Natural Sciences and Engineering
Research Council of Canada; the National Science Council of the Republic of
China; the Swiss National Science Foundation; the A. P. Sloan Foundation; the
Bundesministerium fuer Bildung und Forschung, Germany; the Korea Science
and Engineering Foundation; the Max Kade Foundation; and the Ministry of 
Education, Science and Research of the Federal State Nordrhein-Westfalen 
of Germany.

\setcounter{figure}{0}
\begin{figure}
\centerline{\psfig{figure=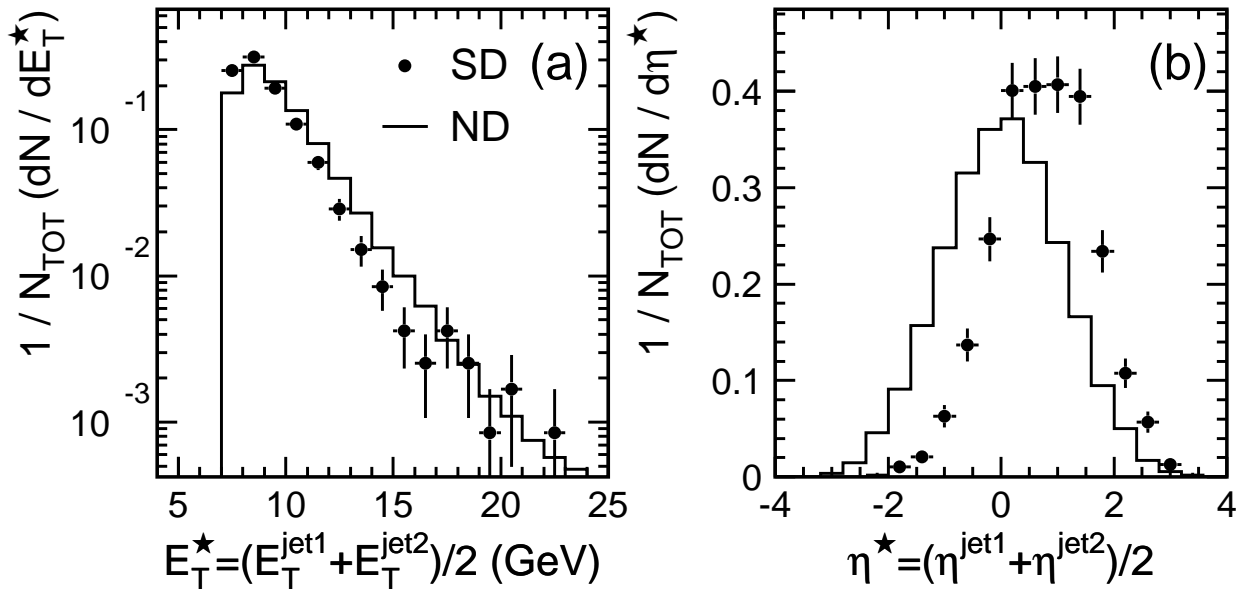,width=7in}}
\vglue 0.25in
\caption{Mean transverse energy and mean pseudorapidity distributions for 
single-diffractive (points) and non-diffractive (histograms) events 
with two jets of $E_T^{jet}>7$~GeV at $\protect\sqrt s=630$ GeV.}
\vglue -2in
\end{figure}
\newpage
\begin{figure}
\vglue 0.5in
\centerline{\psfig{figure=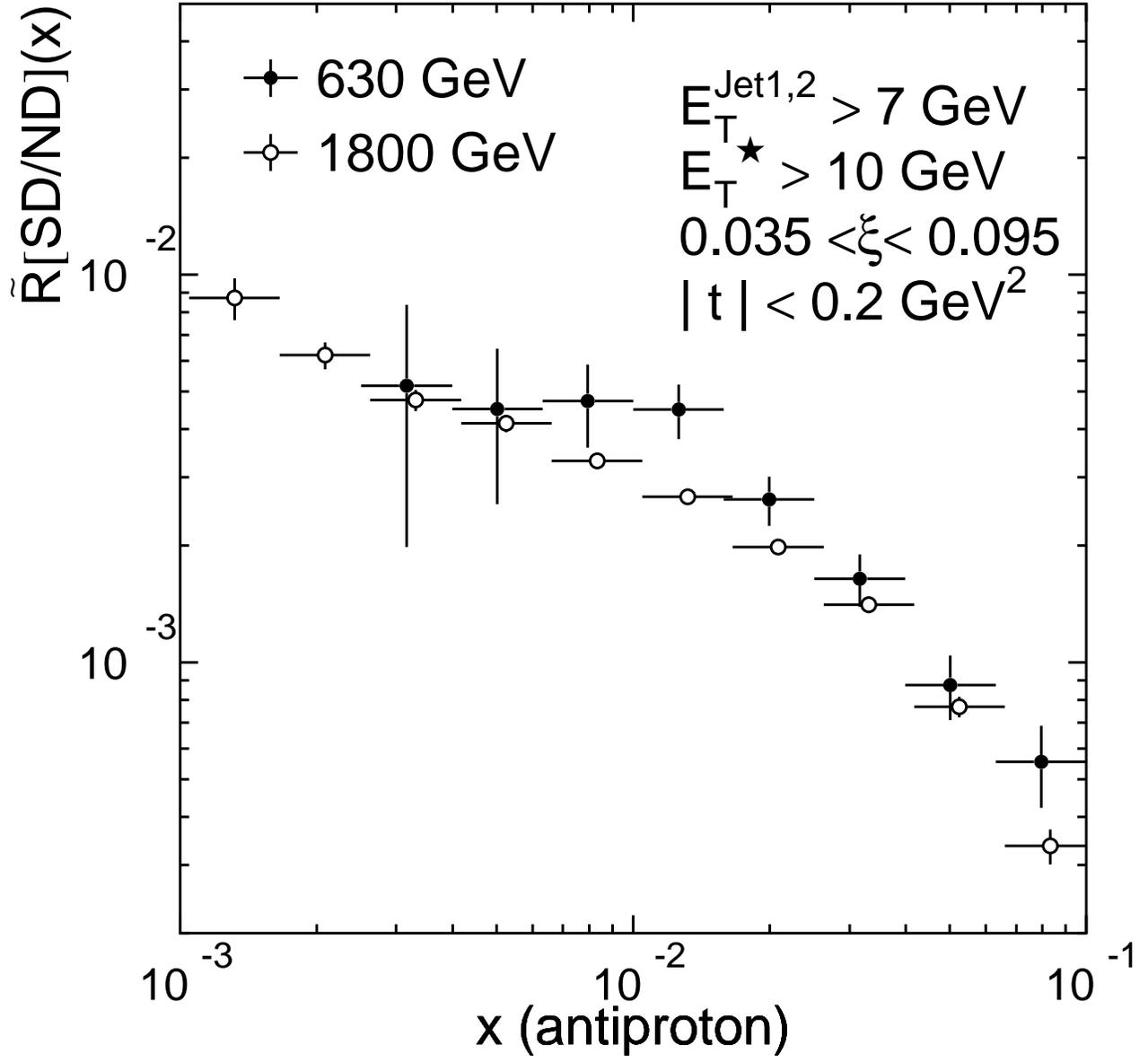,width=7in}}
%\centerline{\psfig{figure=JJ630_PRL_FIG2.EPS,width=3.25in}}
\vfill
\caption{Ratio of single-diffractive to non-diffractive production rates 
as a function of $x$-Bjorken for events with two jets of $E_T>7$ GeV and mean 
$E_T$ greater than 10 GeV at $\protect\sqrt s=630$ GeV (black points) 
and 1800 GeV (open circles). The errors are statistical only.}
\end{figure}
\begin{figure}
\vglue 0.5in
\centerline{\psfig{figure=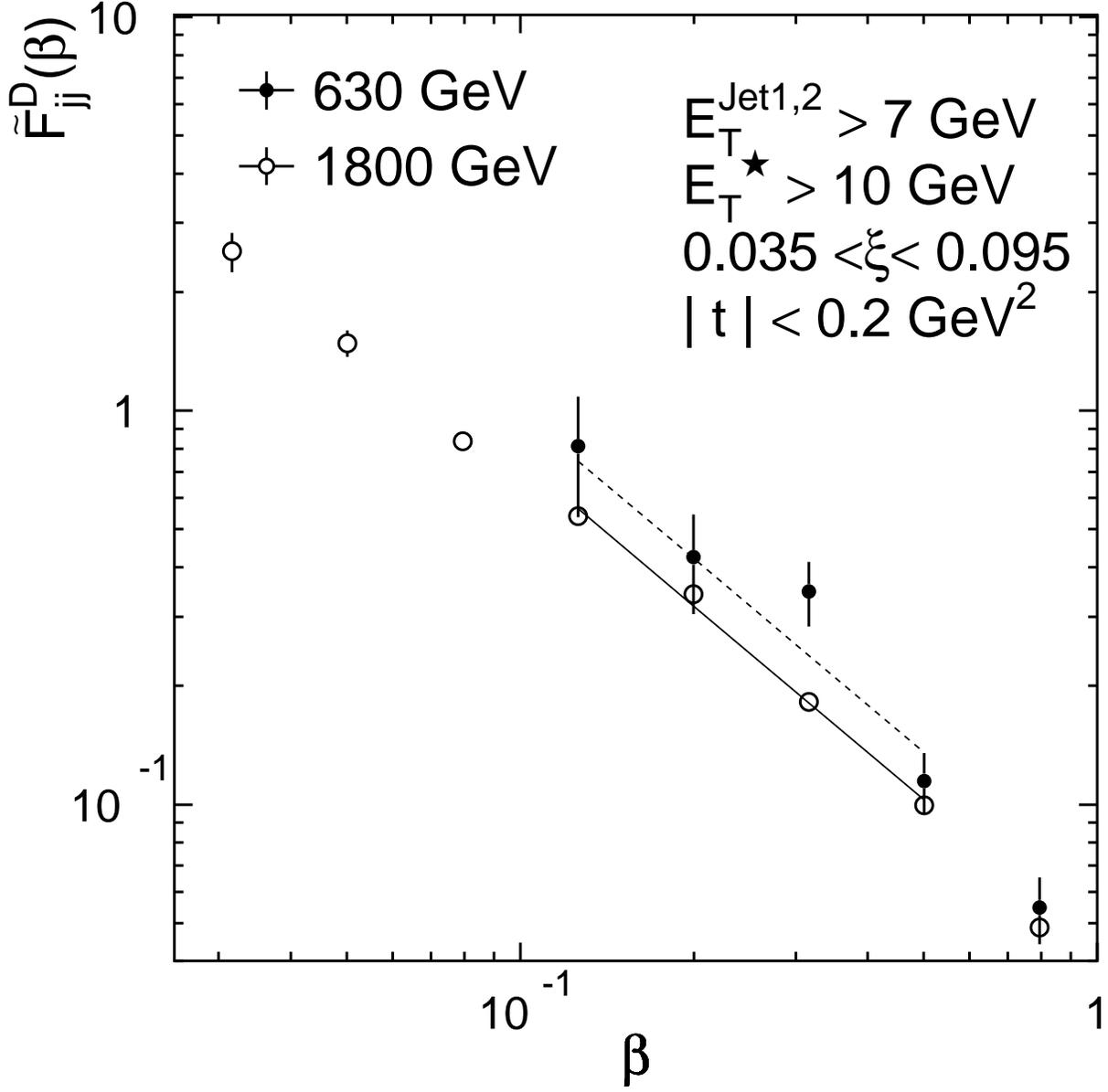,width=7in}}
%\centerline{\psfig{figure=JJ630_PRL_FIG3.EPS,width=3.25in}}
\vfill
\caption{The diffractive structure function versus $\beta$, 
$\tilde{F}^D_{jj}(\beta)$, integrated
over the range $0.035<\xi<0.095$ and $|t|<0.2$ GeV$^2$ and expressed per 
unit $\xi$, at $\protect\sqrt s=630$ GeV (black points) and 1800 GeV 
(open circles). The errors are statistical only.
The lines are fits of the form 
$\beta^{-n}$ with the parameter $n$ common at both energies.
In the fit region, the systematic uncertainty in the ratio of the 630 to 1800
GeV data is $^{+31}_{-23}\%$ (see text).}
\end{figure}
\begin{figure}
\vglue 0.5in
\centerline{\psfig{figure=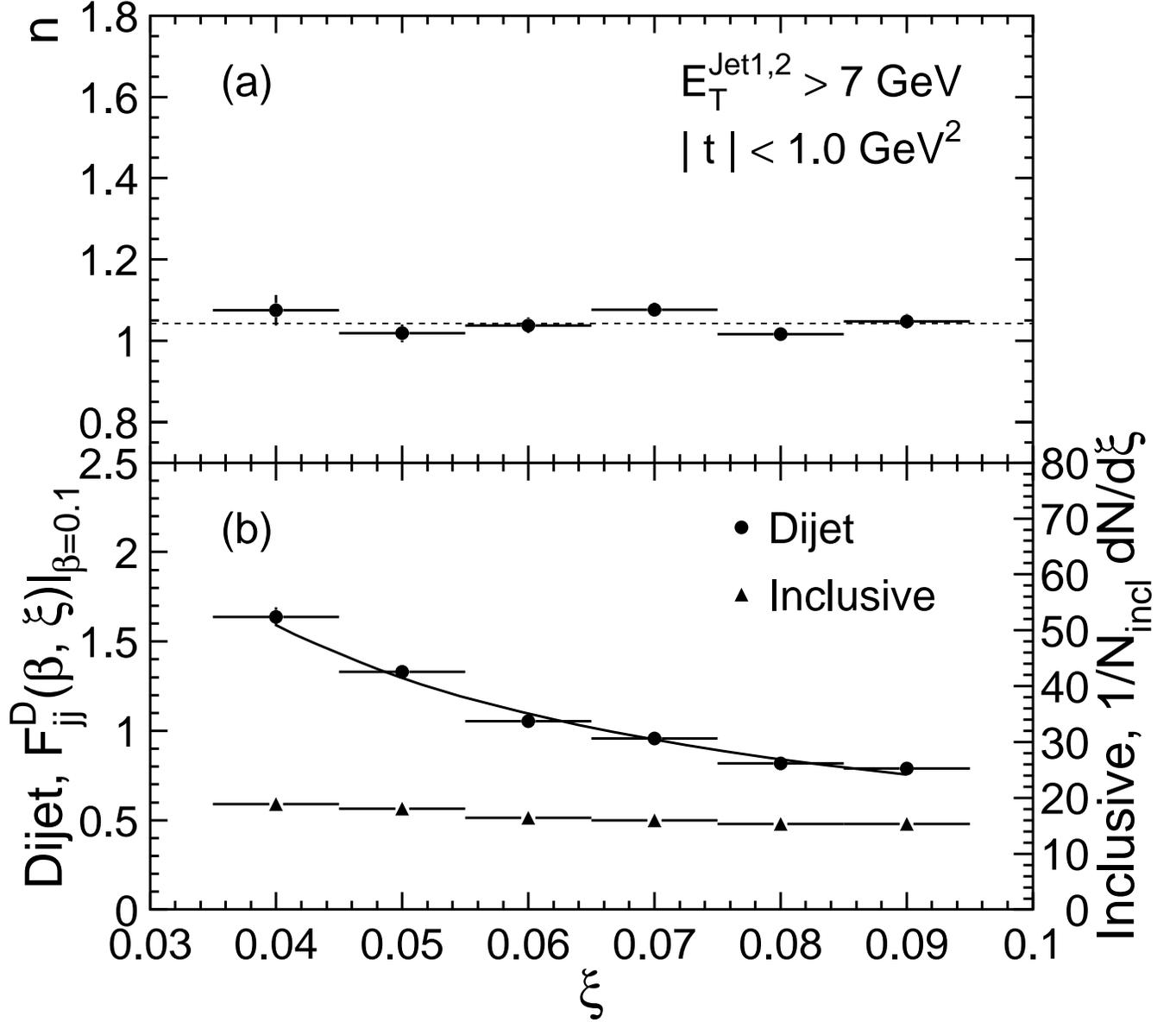,width=7in}}
%\centerline{\psfig{figure=JJ630_PRL_FIG4.EPS,width=3.25in}}
\vfill
\caption{Distributions versus $\xi$ for 1800 GeV data: (a) 
the parameter $n$ of a fit to the diffractive structure function of the form 
$F^D_{jj}(\beta,\xi)|_{\xi}=C\,\beta^{-n}$ for $\beta<0.5$;
(b) the diffractive structure function
at $\beta=0.1$ fitted to the form 
$F^D_{jj}(\beta,\xi)|_{\beta=0.1}=C\,\xi^{-m}$ 
(circle-points and 
curve), and the inclusive single-diffractive distribution (triangles).
The errors shown are statistical.
The fits yield $n=1.0\pm 0.1$ and
$m=0.9\pm 0.1$, where the errors are mainly due to the systematic uncertainties
in the determination of $\beta$.}
\end{figure}
\end{document}